\documentstyle[12pt]{article}
\newcommand{\nc}{\newcommand}

\nc{\dbar}{\bar{\partial}}
\nc{\be}{\begin{equation}}
\nc{\ee}{\end{equation}}

\nc{\beq}{\begin{equation}}
\nc{\eeq}{\end{equation}}
\nc{\bea}{\begin{eqnarray}}
\nc{\eea}{\end{eqnarray}}

\catcode`\@=11

\def\@normalsize{\@setsize\normalsize{15pt}\xiipt\@xiipt
\abovedisplayskip 14pt plus3pt minus3pt%
\belowdisplayskip \abovedisplayskip
\abovedisplayshortskip  \z@ plus3pt%
\belowdisplayshortskip  7pt plus3.5pt minus0pt}
\def\small{\@setsize\small{13.6pt}\xipt\@xipt
\abovedisplayskip 13pt plus3pt minus3pt%
\belowdisplayskip \abovedisplayskip
\abovedisplayshortskip  \z@ plus3pt%
\belowdisplayshortskip  7pt plus3.5pt minus0pt
\def\@listi{\parsep 4.5pt plus 2pt minus 1pt
            \itemsep \parsep
            \topsep 9pt plus 3pt minus 3pt}}

\def\underline#1{\relax\ifmmode\@@underline#1\else
        $\@@underline{\hbox{#1}}$\relax\fi}
\@twosidetrue
\relax

\catcode`@=12

\catcode`\@=11

\def\section{\@startsection{section}{1}{\z@}{3.5ex plus 1ex minus
   .2ex}{2.3ex plus .2ex}{\large\bf}}

\def\ps@headings{\def\@oddfoot{}\def\@evenfoot{}
\def\@oddhead{\hbox{}\hfill
        \makebox[.5\textwidth]{\raggedright\ignorespaces --\thepage{}--
        \hfill }}
\def\@evenhead{\@oddhead}
}

\ps@headings

\catcode`\@=12

\relax

\def\figcap{\section*{Figure Captions\markboth
        {FIGURECAPTIONS}{FIGURECAPTIONS}}\list
        {Fig. \arabic{enumi}:\hfill}{\settowidth\labelwidth{Fig. 999:}
        \leftmargin\labelwidth
        \advance\leftmargin\labelsep\usecounter{enumi}}}
 \relax
\def\tablecap{\section*{Table Captions\markboth
        {TABLECAPTIONS}{TABLECAPTIONS}}\list
        {Table \arabic{enumi}:\hfill}{\settowidth\labelwidth{Table 999:}
        \leftmargin\labelwidth
        \advance\leftmargin\labelsep\usecounter{enumi}}}
 \relax
\def\reflist{\section*{References\markboth
        {REFLIST}{REFLIST}}\list
        {[\arabic{enumi}]\hfill}{\settowidth\labelwidth{[999]}
        \leftmargin\labelwidth
        \advance\leftmargin\labelsep\usecounter{enumi}}}
 \relax

\catcode`\@=11

\def\ps@headings{\def\@oddfoot{}\def\@evenfoot{}
\def\@oddhead{\hbox{}\hfill
        \makebox[.5\textwidth]{\raggedright\ignorespaces --\thepage{}--
        \hfill }}
\def\@evenhead{\@oddhead}
}

\ps@headings

\relax

\def\firstpage#1#2#3#4#5#6{
\begin{document}

\begin{titlepage}
\nopagebreak
\title{\begin{flushright}
       \vspace*{-1.8in}
       {\normalsize UvA-WINS-Wisk. 97-13} 
\end{flushright}
\vfill
{\large \bf #3}}
\author{\large #4 \\ #5}
\maketitle
\vskip -7mm
\nopagebreak
\begin{abstract}
{\noindent #6}
\end{abstract}
\vfill
\begin{flushleft}
\rule{16.1cm}{0.2mm}\\[-3mm]
\end{flushleft}
\footnotesize{PACS: 11.-w } \\
\footnotesize{Keywords: Matrix theory, supergravity, $R^4$-coupling}
\thispagestyle{empty}
\end{titlepage}}
\newcommand{\dal}{\raisebox{0.085cm}
{\fbox{\rule{0cm}{0.07cm}\,}}}
\newcommand{\dt}{\partial_{\langle T\rangle}}
\newcommand{\dtbar}{\partial_{\langle\bar{T}\rangle}}
\newcommand{\al}{\alpha^{\prime}}
\newcommand{\mst}{M_{\scriptscriptstyle \!S}}
\newcommand{\mpl}{M_{\scriptscriptstyle \!P}}
\newcommand{\dv}{\int{\rm d}^4x\sqrt{g}}
\newcommand{\lv}{\left\langle}
\newcommand{\rv}{\right\rangle}
\newcommand{\ph}{\varphi}
\newcommand{\sbar}{\,\bar{\! S}}
\newcommand{\xbar}{\,\bar{\! X}}
\newcommand{\fbar}{\,\bar{\! F}}
\newcommand{\zbar}{\,\bar{\! Z}}
\newcommand{\tbar}{\bar{T}}
\newcommand{\ubar}{\bar{U}}
\newcommand{\ybar}{\bar{Y}}
\newcommand{\phb}{\bar{\varphi}}
\newcommand{\cm}{Commun.\ Math.\ Phys.~}
\newcommand{\pr}{Phys.\ Rev.\ D~}
\newcommand{\prl}{Phys.\ Rev.\ Lett.~}
\newcommand{\pl}{Phys.\ Lett.\ B~}
\newcommand{\ibar}{\bar{\imath}}
\newcommand{\jbar}{\bar{\jmath}}
\newcommand{\np}{Nucl.\ Phys.\ B~}
\newcommand{\e}{{\rm e}}
\newcommand{\gsi}{\,\raisebox{-0.13cm}{$\stackrel{\textstyle
>}{\textstyle\sim}$}\,}
\newcommand{\lsi}{\,\raisebox{-0.13cm}{$\stackrel{\textstyle
<}{\textstyle\sim}$}\,}
\date{}
\firstpage{95/XX}{3122}
{\large\sc\bf A comment on the $R^4$-coupling in (M)atrix Theory} 
{Marco Serone}
{\normalsize\sl Dept. of Mathematics, University of Amsterdam \\[-3mm]
\normalsize\sl Plantage Muidergracht 24, 1018 TV Amsterdam, The Netherlands\\[-3mm]
\normalsize\sl e-mail: serone@wins.uva.nl \\[-3mm]}
{By scaling arguments we show that the presence of a $R^4$-term in the eleven 
dimensional supergravity effective lagrangian, if it is visible in (M)atrix theory,
should produce a correction to the five-loop effective lagrangian of two
moving D0-branes.}
\newpage 
There is no doubt that during last year there has been an increasing interest about the
conjecture of Banks, Fischler, Shenker and Susskind (BFSS) \cite{bfss}. They basically consider
a non-perturbative formulation of M-theory in a fixed kinematical regime as given by the
effective Super Yang-Mills $U(N)$ quantum mechanics \cite{nqm} governing the IR dynamics
of a system of $N$ D0-branes. More precisely the original prescription stated that the
D0-brane two-derivative lagrangian, parametrized by the string coupling constant $g_S$ and
the string length $l_S$, is a non-perturbative description of M-theory
compactified on a spatial circle of radius $R$ and Planck length $l_P$ with the
identifications
\be R=g_S\,l_S, \ \ \ \ l_{P}=g_S^{1/3}\,l_S \label{eq0} \ee
This correspondence is realized in the infinite momentum frame where for any state with 
positive momentum $P$ along $R$,
\be P=\frac{N}{R}\rightarrow\infty, \ \ \ N\rightarrow\infty \ee
Later on, Susskind \cite{suss} realized that Matrix Theory (MT) is meaningful even for $N$
finite, if we consider it as M-theory on a {\it light-like} circle. In this way, he showed
that MT at finite $N$ is just the so-called Discrete Light Cone Quantization of M-theory.
Up to now a lot of tests have been performed to the BFSS conjecture 
\cite{test1,bc,pp}\footnote{We cite here only those works that provided evidence of MT through
some explicit scattering computations. See \cite{ban} and references therein for a recent
and wider perspective of MT.}, some of
them performed at finite $N$ \cite{test2,bbpt,ct}, using Susskind's prescription. For the
simplest case of compactifications on a circle, all of them showed an agreement between the MT
and the Sugra computation of same physical processes \footnote{Note, however, that very
recently a discrepancy for multigraviton scattering \cite{dr} appeared.}.
The case of compactified MT, and in particular to non-trivial spaces, is more involved and
present some problems \cite{comp}.\\
Very recently Seiberg \cite{sei} (see also \cite{sen}) gave a prescription that helps
to clarify the range of validity of MT and gives new evidence to the point of view of
\cite{suss}. In few words, he derived Susskind's conjecture, relating M-theory
compactified on a light-like circle to another M-theory on a spatial circle. More
precisely, his prescription implies that
any amplitude ${\cal A}_{D0}(g_S,l_S)$ computed starting from the
D0-brane lagrangian is related to the corresponding amplitude 
${\cal A}_M(l_P,R)$ of the M-theory on the spatial circle of radius $R$ and Planck length $l_P$ 
in the following limits:
\bea \lim_{g_S,\,l_s\rightarrow 0}{\cal A}_{D0}\,(g_S,l_S)&=&
\lim_{R,\,l_P\rightarrow 0} {\cal A}_M\,(R,l_P) \label{lim} \\ 
{\rm with} \ \ \  l_S\,g_S^{-1/3}={\rm fixed}, 
&& \ \ l_P^2\,R^{-1}={\rm fixed} \nonumber \eea
where $N$ is again related to the momentum $P$ as $P=N/R$. First of all, note that there is
not anymore any $N$-limit in the correspondence, that is then valid for any (positive) value
of $N$. The important point is that the limit on the left side of eq.(\ref{lim}) defines
precisely the perturbative regime of MT, on the contrary of the original proposal, where an
extrapolation at strong coupling of amplitudes was needed.\\
Despite its success, MT has not been up to now a source of predictions; it has been used 
instead to find in new ways results already known in string or field theory. MT, however,
is supposed to be a non-perturbative formulation of M-theory and then, by definition, a 
microscopic theory underlying eleven dimensional supergravity. In this respect, it should 
remove the bad UV-divergences that plague 11D Sugra, in much the same way 
string theory does. MT should be able, for instance, to fix the coefficient
(otherwise divergent) multiplying (one of) the $R^4$-term appearing in the 11D Sugra
action \cite{gv,ggv}; even if this finite coefficient can be fixed by requiring consistency 
with string theory, an explicit MT result should be seen as an alternative and more
direct procedure and, at the same time, as a strong check to the validity of the theory
itself. \\
Aim of this note is then to show that, if this term is in anyway visible in MT, it should
appear as a correction to the effective lagrangian of two moving D0-branes induced by a 
{\it non-planar diagram} at five loops.

We will see, moreover, that according to the correspondence (\ref{lim}), the $N$-finite
MT before the $g_S,l_S\rightarrow 0$ limits, can be seen as a ``regularized version'' of eleven
dimensional supergravity (compactified on a circle);
in this way, among the infinite series of terms appearing at any loop 
$l$ in MT for the potential between two D0-branes, all the terms proportional to
$v^n$, with $n>2l+2$, are simply an effect of the regularization and, after an appropriate
rescaling, vanish in the limit (\ref{lim}). 
In this perspective, on the other hand the
``would-be'' 5-loop term, responsible in MT of the $R^4$-coupling, diverges in the limit
(\ref{lim}), reproducing the UV-divergence occuring in Sugra for this term. 

Let us then consider the effective lagrangian for two D0-branes moving with relative velocity
$v$ and separation $r$; by dimensional analysis it is straightforward to see that at loop $l$,
their effective lagrangian $L^{(l)}$ can be written as \cite{eff}:
\be L^{(l)}=g_S^{l-1}\,\sum_{n=0}^{\infty}C_{nl}\,v^2\,
\left(\frac{v^2}{r^4}\right)^n
\left(\frac{1}{r^3}\right)^l\,l_S^{4n+3l-1} \label{eq2} \ee
where the powers of $l_S$ are fixed requiring $L^{(l)}$ to have the dimension of a 
$({\rm length})^{-1}$ and we omitted the $N$-dependence that will be discussed later. 
It is known that at 1-loop, $L^{(1)}\sim v^4/r^7$ ($C_{01}=0$),
whereas the two-loop effective lagrangian starts as $v^6/r^{14}$ \cite{bbpt}
($C_{02}=C_{12}=0$). It has been conjectured by \cite{ct} that at $l^{th}$ 
loop order, $C_{0l}=...=C_{l-1,l}=0$, in order to reproduce classical long-distance
supergravity potentials. We will see that from the supergravity point of view 
this assertion is consistent, provided that we neglect higher derivative operators in the 
low-energy effective action\footnote{This observation has been also independently done in
a recent paper \cite{alw}.}.
Consider indeed the interaction of two massless particles (in 11D), considered as external 
sources, with gravity. The corresponding action is simply  \footnote{For what follows we 
will need simply the Einstein term of the Sugra action.}:
\be S=-\frac{1}{2k_{11}^2}\int\! d^{11}\!x\,\sqrt{g}\,R+
\int\!d^{11}\!x\,\sqrt{g}\,g_{\mu\nu}\,T^{\mu\nu}
\label{eq3} \ee
where it is understood that one direction (the $11^{th}$) is compactified on a circle of
radius $R_{11}$, the gravitational constant $k_{11}^2=16\pi^5 l_P^9$ \cite{bc} and
\be T^{\mu\nu}(x)=\sum_{i=1,2}\frac{P^{\mu}_iP^{\nu}_i}{E_i}\,
\delta^{(10)}(x^{\mu}-x^{\mu}_i(t)) \label{tmn} \ee
We choose 
\bea x_1^{\mu}(t)&=&(t,\frac{v}{2}t,\frac{b}{2},0,...,0,v_{11}t) \label{x} \\
         x_2^{\mu}(t)&=&(t,-\frac{v}{2}t,-\frac{b}{2},0,...,0,v_{11}t) \nonumber \eea
with $v<<v_{11}\sim 1$, with the corresponding momenta $P^{\mu}_i$ ($P^{11}=N/R_{11}$). 
According to eq.(\ref{lim}), it is meaningful to consider a loopwise expansion, with
$k_{11}^2$ as Planck constant. Expand then eq.(\ref{eq3}) around the flat metric 
$g_{\mu\nu}=\eta_{\mu\nu}+k_{11}\,h_{\mu\nu}$ and take the graviton propagator
in the De Donder gauge:
\be \Delta^{\mu\nu,\rho\sigma}(x-y)=(\eta^{\mu\rho} \eta^{\nu\sigma}+
\eta^{\mu\sigma} \eta^{\nu\rho} -\frac{2}{9}\eta^{\mu\nu} \eta^{\rho\sigma})
\frac{1}{2\pi R_{11}}\int\!\frac{d^{10}\!q}{(2\pi)^{10}}\,\frac{e^{iq(x-y)}}{q^2} \label{don} \ee 
where we have fixed $q_{11}=0$.
It is then an easy exercise, using eqs.(\ref{tmn}), (\ref{x}) and (\ref{don}),  to obtain the 
familiar leading $v^4/(b^2+v^2t^2)^{7/2}$ dependence of the potential between the two
sources:  a simple algebra shows that the tensorial structure of eqs.(\ref{tmn}) and (\ref{don})
imply that each source carries effectively a power of $P_{11}\cdot v^2$, whereas the graviton 
propagator gives a factor $\sim 1/R_{11}r^7$; these are of course the leading terms in an
expansion in $v<<1$. Changing to the string units $g_S,l_S$, 
we have exactly the same scaling provided by eq.(\ref{eq2}). Note, however, that if we rescale both
amplitudes and then take the limit (\ref{lim}), all the terms $C_{n1}, n>1$ in eq.(\ref{eq2})
vanish, as previously mentioned. \\
In this kind of expansion in $k_{11}$, all higher order corrections 
scale necessarily like $(k_{11}^2)^m\sim l_P^{9m}$, with $m$ an integer. This condition is
indeed satisfied by eq.(\ref{eq2}), provided $l=n$. For instance, the next to leading
correction to the potential is given by contracting three source terms with the term trilinear 
in $h_{\mu\nu}$, coming from the expansion of the scalar curvature $R$. 
Even if the full contribution is rather involved, it is easily seen that the end 
result goes as $v^6/r^{14}$. 
The next order, coming from the four-graviton term in $R$, scales like
\be V_R\sim \frac{l_P^{27}v^8P_{11}^4}{R^3_{11}r^{21}} \label{r} \ee
where the 9-dimensional integration over the position of the interaction vertex cancels a 
contribution of one of the four propagators and the two derivatives of the vertex itself,
leading then to eq.(9).  In MT this correction should be visible as the three-loop effect 
$l=n=3$ in eq.(\ref{eq2}). It is clear that these terms correspond to the classical expansion 
of the potential found in \cite{bbpt} with a source-probe analysis.
It has been conjectured by \cite{gv} that the 11D Sugra action should contain a $R^4$-term,
that is needed to reproduce some perturbative and non-perturbative terms in IIA,B string
theory. For M-theory compactified on a circle $R_{11}$, the analysis of \cite{gv} shows 
the presence of two contributions for $R^4$: 
\be  S_{R^4}=\frac{R_{11}}{k_{11}^{2/3}}\int\! d^{10}\!x\,\sqrt{g}\,t_8t_8\,R^4
\left(\frac{2\pi^2}{3}+2\zeta (3)\frac{l_P^3}{R^3_{11}}\right) \label{r4s} \ee
up to an overall factor, where the tensorial structure $t_8t_8\,R^4$ is explicitly given in \cite{tse}.
For the purpose of the present analysis, we will not need the detailed form of this
term; from the Sugra point of view, these couplings arise as counterterms coming from a
one-loop four graviton scattering. The first term in eq.(\ref{r4s}) is actually
UV-divergent and its finite coefficient is fixed by consistency with IIA,B string
theory and T-duality, whereas the second one is completely finite \cite{ggv}. 
We will see that the second term in eq.(\ref{r4s}) is not visible in our MT computation
(it does not match with our scaling arguments);
we will focus our attention in the following on the first term only. 

The presence of the $R^4$-coupling (\ref{r4s}) to the effective Sugra action produces a 
sub-leading correction
to the effective potential of the sources; in particular its linear structure contains a
four-graviton term that induces a correction proportional to $v^8/r^{27}$, where the additional
six powers of $r$ with respect to eq.(\ref{r}) come from the six more derivatives contained in 
$R^4$, compared to the four-graviton two-derivative term of the $R$-coupling.
From eq.(\ref{eq2}), we see that a term of this form in MT is possible for $l=5,n=3$ and would be
an ``anomaly'' to the suggestion that $C_{0l}=...=C_{l-1,l}=0$; this is however expected since 
the correction (\ref{r4s}) presents a fractional power of the gravitational constant $k_{11}$. 
We will see that this is consistent with our scaling arguments; the correction induced by the 
(first) term in eq.(\ref{r4s}) is
\be V_{R^4}^{Sugra}=c_S\,l_P^{33}\frac{v^8}{r^{27}}\,\frac{P_{11}^4}{R^3_{11}} 
\label{eq5} \ee
where $c_S$ is a numerical coefficient. The corresponding MT potential is
\be V_{R^4}^{MT}=c_M\,l_S^{26}\,g_S^4\frac{v^8}{r^{27}} \label{eq6} \ee
Using eq.(\ref{eq0}), we see that there is a complete matching between the scaling
of both terms,
\be \frac{V_{R^4}^{Sugra}}{V_{R^4}^{MT}}=\frac{c_S}{c_M} \label{eq7}\ee
where we omitted the $N$-dependence in both eqs.(\ref{eq6}) and (\ref{eq7}), dependence that 
will be discussed in the following.
The presence of a non-vanishing coefficient $c_M$ in eq.(\ref{eq6}) 
would be a clear signal that MT is in the right way.
If we normalize to a finite value in MT the ``standard'' $v^{2(l+1)}$-term,
then the limit (\ref{lim}) gives a divergent result for $V_{R^4}^{MT}$, that indeed reproduces
the UV-divergence of this term in Sugra. 
The $g_S$-dependence of 
eq.(\ref{eq6}) and then the origin of five-loops is easily seen in string theory: it is known
that a generic two-dimensional surface with $g$ handles, $c$ crosscaps and $b$ boundaries 
has a string coupling constant dependence given by $g_S^{(2g-2+b+c)}$. The usual annulus 
computation ($g=c=0,b=2$) gives of course $g_S^0$, whereas the two and three loop MT 
amplitudes responsible for corrections of order $v^6/r^{14},v^8/r^{21}$ respectively, 
correspond to a sphere with 3,4 punctures ($g=c=0,b=3,4$)\footnote{Each puncture corresponds,
from the Sugra point of view, to an insertion of the source term given by eq.(\ref{eq3}).}. 
The $R^4$-coupling
we have considered is just the term arising at
one-loop in IIA string theory \cite{gv}, as can be easily seen using again eq.(\ref{eq0}); 
the corresponding surface is then a torus with four 
punctures ($g=1,c=0,b=4$) giving a $g_S^4$ dependence, coming in MT precisely at
five loops. Let us consider now the $N$-dependence of these amplitudes; 
the leading large $N$-behaviour of eq.(\ref{eq2}) at loop $l$ 
is given by $N^{l+1}$, that does not match at five loops with the Sugra result 
$\sim N^4$. This is indeed what we expect since we have just seen that the two-dimensional
surface responsible of this interaction in string theory involves a genus one surface,
that leads then to a {\it non-planar diagram} in the Super Yang-Mills quantum mechanics,
and with a sub-leading $N$ power-dependence. Since every boundary on the world-sheet involves
a trace on the fundamental of the gauge group, each of them gives one power of $N$, that 
then matches precisely with the analysis above.\\
A comment on the $2^{nd}$ term of eq.(\ref{r4s}) is however needed.
As already mentioned, the correction induced by this term does not match with the MT
parameters; it should be visible as a planar diagram at three-loops, according with the 
analysis above, since it arises at tree-level in IIA string theory; at this order, however,
the only $v^8$-term appearing is that induced by the four-graviton term of the scalar 
curvature $R$, as in eq.(\ref{r}). This mis-matching could be due to the fact that
this term arises probably as a counterterm generated by integrating out Kaluza-Klein states
with non-vanishing momentum around $R_{11}$ and is then not visible in a MT computation 
where we do not allow exchange of momentum in the eleventh direction. \\
This discussion can be also extended to more general higher derivative terms;
any non-vanishing term in, say, the effective potential of two moving D0-branes
with $n<l$ in eq.(\ref{eq2}), correspond to a counterterm of higher dimension in the Sugra action
(see also \cite{alw} for a similar remark).
In this perspective MT could in principle predict the form of new couplings and
the finite correct values of their coefficients, in general divergent in Sugra,
altough they should come out from very laborious many-loops computations in the Super Yang-Mills
quantum mechanics.

Finally, I would like to comment on other papers \cite{eff,kk}, where it is shown that the
presence of the $R^4$-coupling in the supergravity effective action should give corrections
to the potential of two gravitons going like $v^8/r^{18}$. In \cite{eff} it is argued that the
N-dependence of this term is consistent with the two loop term ($l=2,n=3$) of eq.(4),
whereas the analysis of \cite{kk} shows a disagreement in the powers of $N$, not
reconciliable with MT\footnote{Note, however, that in a  recent paper \cite{bgl}  a possible 
explanation to the N-scaling of \cite{kk} is given.}. The reason of the apparent conflict
between the results of \cite{eff,kk} with those presented here, is due to the different
approaches followed. As shown before, we treat the external states in sugra as classical 
spinless sources, corresponding to the MT background of two moving D0-branes. The limit
we consider corresponds precisely to weakly coupled IIA string theory in 10 dimensions
with two clusters of coincident D0-branes. On the other hand refs.\cite{eff,kk} consider
the external states as pure (super-)gravitons, with definite spin. The $v^8/r^{18}$ term
above is indeed found by considering also graviton polarizations; this effect is then not 
visible in our approximation where the external states are spinless. On the other hand, the
source-probe analysis of \cite{eff,kk} cannot capture the term considered in this paper,
being a non-local term in the effective action for the probe\footnote{I thank E. Keski-Vakkuri and
P. Kraus for a correspondence that helped me to clarify these points.}.

It is not excluded, of course, the simplest possibility that $c_M$ in eq.(\ref{eq6}) is zero,
just for kinematical reasons.
It has been shown in \cite{bbpt}, for instance, that the gravitational field created by a massless 
graviton in the light-cone frame cannot receive any correction by possible $R^n$-terms present in the
lagrangian, because they are all vanishing in this frame \cite{hs}. I do not exclude that this is
indeed what happens; a more precise computation in Sugra should answer this question.\\
Needless to say, it would be of the utmost importance to understand how and where to find
in MT compactified on a circle and on a torus the results of \cite{gv,ggv} for $R^4$ in nine and
eight dimensions. I believe that a clear understanding of these (and similar)
higher derivative interactions will shed a new light on the range of validity
of Matrix Theory as a non-perturbative formulation of M-theory.  

{\bf{Acknowledgements}}

It is a pleasure to thank R. Dijkgraaf, E. Verlinde and H. Verlinde for useful and 
enlightening discussions.



\begin{thebibliography}{99}
\bibitem{bfss} T. Banks, W. Fischler, S. Shenker and L. Susskind, {\it
M-theory as a matrix model: a conjecture}, Phys. Rev. {\bf D55} (1997) 112,
hep-th/9610043;
\bibitem{nqm} M. Claudson and M. Halpern, {\it Supersymmetric ground state wave
functions}, Nucl. Phys. {\bf B250} (1985) 689; \\
M. Baake, M. Reinicke and V. Rittenberg, {\it Fierz identities for real Clifford
algebras and the number of supercharges}, J. Math. Phys. {\bf 26} (1985) 1070;\\
R. Flume, {\it On quantum mechanics with extended supersymmetry and non abelian gauge
constraints}, Ann. Phys. {\bf 164} (1985) 189;
\bibitem{suss} L. Susskind, {\it An other conjecture about (M)atrix Theory},
hep-th/9704080;
\bibitem{test1} O. Aharony and M. Berkooz, {\it Membrane dynamics in (M)atrix Theory},
Nucl. Phys. {\bf B491} (1997) 184, hep-th/9611215;\\
G. Lifschytz and S.D. Mathur, {\it Supersymmetry and membrane interactions in (M)atrix
Theory}, hep-th/9612087;\\
G. Lifschytz, {\it Four-brane and six-brane interactions in (M)atrix Theory},
hep-th/9612223; {\it A note on the transverse five-brane in (M)atrix Theory},
hep-th/9703201;\\
I. Chepelev and A.A Tseytlin, {\it Long-distance interactions of D-brane bound states
and longitudinal 5-brane in (M)atrix theory}, Phys. Rev. {\bf D56} (1997) 3672,
hep-th/9704127;\\
J.M. Pierre, {\it Interactions of eight-branes in string theory and (M)atrix Theory},
hep-th/9705110;\\
R. Gopakumar and S. Ramgoolam, {\it Scattering of zero branes off elementary strings
in Matrix Theory}, hep-th/9708022;
\bibitem{bc} D. Berenstein and R. Corrado, {\it M(atrix) Theory in various dimensions},
Phys. Lett. {\bf B406} (1997) 37, hep-th/9702108;
\bibitem{pp} J. Polchinski and P. Pouliot, {\it Membrane scattering with M-momentum 
transfer}, hep-th/9704029;
\bibitem{ban} T. Banks, {\it Matrix Theory}, hep-th/9710231;
\bibitem{test2} K. Becker and M. Becker, {\it A two loop test of (M)atrix Theory},
hep-th/9705091;\\
P. Kraus, {\it Spin-orbit interaction from Matrix Theory}, hep-th/9709199;
\bibitem{bbpt} K. Becker, M. Becker, J. Polchinski and A.A. Tseytlin, {\it Higher order 
graviton scattering in (M)atrix Theory}, Phys. Rev. {\bf D56} (1997) 3174,
hep-th/9706072;
\bibitem{ct} I. Chepelev and A.A. Tseytlin, {\it Long-distance interactions of branes:
correspondence between supergravity and super Yang-Mills descriptions}, hep-th/9709087;\\
E. Keski-Vakkuri and P. Kraus, {\it Born-Infeld actions from Matrix Theory}, hep-th/9709122;
\bibitem{dr} M. Dine and A. Rajaraman, {\it Multigraviton scattering in the Matrix
Model}, hep-th/9710174;
\bibitem{comp} M.R. Douglas, H. Ooguri and S.H. Shenker, {\it Issues in (M)atrix Theory
compactification}, Phys. Lett. {\bf B402} (1997), hep-th/9702203;\\
O.J. Ganor, R. Gopakumar and S. Ramgoolam, {\it Higher loop effects in M(atrix)
orbifolds}, hep-th/9705188;\\
M.R. Douglas, {\it D-branes and Matrix Theory in curved space}, talk given at Strings
'97, hep-th/9707228;\\
M.R. Douglas and H. Ooguri, {\it Why Matrix Theory is hard}, hep-th/9710178;
\bibitem{sei} N. Seiberg, {\it Why is the Matrix Model correct ?}, hep-th/9710009;
\bibitem{sen} A. Sen, {\it D0 branes on $T^n$ and Matrix Theory}, hep-th/9709220; 
\bibitem{gv} M.B. Green and P. Vanhove, {\it D-instantons, strings and M-theory},
hep-th/9704145;
\bibitem{ggv} M.B. Green, M. Gutperle and P. Vanhove, {\it One loop in eleven
dimensions}, hep-th/9706175;
\bibitem{eff} P. Berglund and D. Minic, {\it A note on effective lagrangians in Matrix
Theory}, hep-th/9708063;
\bibitem{alw} S.P. de Alwis, {\it Matrix models and string world sheet duality}, 
hep-th/9710219;
\bibitem{tse} A.A. Tseytlin, {\it Heterotic-type I superstring duality and low-energy
effective actions}, Nucl. Phys. {\bf B467} (1996) 383, hep-th/9512081;
\bibitem{kk} E. Keski-Vakkuri and P. Kraus, {\it Short distance contributions to graviton-graviton
scattering: matrix theory versus supergravity}, hep-th/9712013;
\bibitem{bgl} V. Balasubramanian, R. Gopakumar and F. Larsen, {\it Gauge theory, geometry and the
large N limit}, hep-th/9712077;
\bibitem{hs} G.T. Horowitz and A.R. Steif, {\it Spacetime singularities in string
theory}, Phys. Rev. Lett. {\bf 64} (1990) 260.
\end{thebibliography}
\end{document}